\documentclass[structabstract]{aa} 
\usepackage{graphicx}
\usepackage{txfonts}
\begin{document}
  \title{A mass threshold in the number density of passive galaxies at z$\sim$2}


  \author{V. Sommariva
   \inst{1}
     \and A. Fontana \inst{1}
      \and A. Lamastra \inst{1}
            \and P. Santini \inst{1}
        \and J. S. Dunlop \inst{3}
          \and M. Nonino \inst{2}
        \and M. Castellano \inst{1}
      \and H. Ferguson \inst{4}
      \and R. J. McLure \inst{3}
     \and A. Galametz \inst{1}      
           \and M. Giavalisco \inst{5}
       \and A. Grazian \inst{1}
               \and Y. Lu \inst{6}
        \and N. Menci  \inst{1}
        \and A. Merson \inst{7}
        \and D. Paris \inst{1}
         \and L. Pentericci \inst{1}
  	 \and R. Somerville\inst{4}
	     \and T. Targett \inst{3}
	      }
    \institute{INAF - Osservatorio Astronomico di Roma, Via Frascati 33, IÐ00040, Monteporzio, Italy
      \email{veronica.sommariva@oa-roma.inaf.it}
      \and
         INAF - Osservatorio Astronomico di Trieste, Via G.B. Tiepolo 11, 34131 Trieste, Italy.
                   \and
     SUPA, Institute for Astronomy, University of Edinburgh, Royal Observatory, Edinburgh, EH9 3HJ, UK   
      \and
          Space Telescope Science Institute, 3700 San Martin Dr., Baltimore, MD 21218, USA
          \and
      Astronomy Department, University of Massachusetts, Amherst, MA 01003, USA
          \and 
            kavli institute for particle astrophysics and cosmology, 452 Lomita Mall Stanford, CA  94305-4085
            \and
            Department of Physics and Astronomy, University College London, Gower Street, London WC1E 6BT, UK
	}

  \date{Received ; accepted}

 
 \abstract
 {The process that quenched  star formation in galaxies at intermediate and high redshift is still the subject 
of considerable debate.
One way to investigate this puzzling issue is to study the number density of quiescent
galaxies at $z \simeq 2$, and its dependence on mass. Here we present the results of a new study 
based on very deep $K_s$-band imaging (with the HAWK-I instrument on the VLT) of 
two HST CANDELS fields (the UKIDSS Ultra-deep survey (UDS) field and GOODS-South).
The new HAWK-I data (taken as part of the HUGS VLT Large Program) reach detection limits of 
$K_s > 26$ (AB mag). We 
select a sample of passively-evolving galaxies in the redshift range 
$1.4<z<2.5$ (via the $pBzK$ color-based selection criterion of Daddi et al. 2004). 
Thanks to the depth and large area coverage of our imaging, we have been able to extend the selection of 
quiescent galaxies a magnitude fainter than previous analyses. 
Through extensive simulations we demonstrate, for the first time, that the observed turn-over in the 
number of quiescent galaxies at $K\geq 22$ is not due to incompleteness, but is real. 
This has enabled us to establish unambiguously that the number counts of quiescent 
galaxies at $z \simeq 2$ flatten and slightly decline at magnitudes fainter than $K_{s} \sim 22$ (AB mag.).
We show that this trend corresponds to a stellar mass threshold $M_* \simeq 10^{10.8}\,{\rm M_{\odot}}$ below which 
the mechanism that halts the star formation in high-redshift galaxies seems to be inefficient. 
We also show that
at $K>23$ a  higher redshift
population of $z\simeq 3$ $pBzK$ galaxies is detected, and dominates the counts at the faintest magnitudes.
Finally we compare the observed $pBzK$ number counts with those of quiescent galaxies extracted 
from four different semi-analytic models. We find that only two of these models 
reproduce even qualitatively the observed trend in the number counts, and that none of the models 
provides a statistically acceptable description of the number density of quiescent galaxies at these 
redshifts. We conclude that the mass function of quiescent galaxies as a function of redshift continues
to present a key and demanding challenge for proposed models of galaxy formation and evolution.
}
 

  \keywords{Galaxies: high redshift - Galaxies: fundamental parameters -
}

 \maketitle
%

\section{Introduction}

The cessation, or ``quenching'' of star formation activity 
in galaxies at high and intermediate redshifts is one of the 
key events in the history of galaxy evolution, and one 
that is not easily explained in the context of hierarchical 
growth within a $\Lambda$CDM cosmology. While it is now established 
that the (apparently fairly rapid) termination of star formation 
in some subset of the galaxy population only a few billion years 
after the Big Bang must be attributed to some aspect of baryonic 
physics, the relevant process remains a matter of debate,
with opinion divided over whether the key mechanism is feedback from 
an active galactic nucleus (AGN), feedback from star formation itself,
or termination of the gaseous fuel supply (e.g. due to the difficulty
of cool gas penetrating to the centre of a massive halo due to shock 
heating). Better observational constraints on the prevalence 
and properties of quiescent galaxies at high redshift are 
therefore urgently required in order
to constrain models of galaxy formation and evolution over cosmic 
history. 

One key goal is to better establish the mass dependence of star-formation 
quenching. A number of lines of evidence suggest that this process appears
``anti-hierarchical'' in the sense that it is the most massive galaxies 
that cease star formation at the earliest times, leading ultimately 
to the present-day Universe in which the most massive elliptical galaxies
appear to be the oldest. However, because less-massive passive galaxies 
are hard to detect and isolate at high redshift (especially at optical wavelengths),
it has proved difficult to establish the definitive mass dependence of 
the number density of passive galaxies at the crucial epoch corresponding to
$z \simeq 2$. The primary aim of this study is to clarify this situation 
via new deep multi-frequency imaging data.

Several methods have been developed to efficiently select passively-evolving 
galaxies at high-redshift based on color criteria.
In recent years, the most extensively used selection technique is the 
$BzK$ color-color selection introduced by Daddi et al. (2004). They showed that galaxies 
in the redshift range $1.4<z<2.5$ fall into a specific area of the $(B-z)$ / $(z-K)$ diagram, and demonstrated  
that galaxies can be easily separated into star-forming ($sBzK$) and passively-evolving 
($pBzK$) classes. This technique also has the advantage of not being biased by the presence of 
dust; 
the reddening vector in the $BzK$ plane 
is parallel to the $BzK$ selection line, making this criterion relatively immune to dust content. 

A number of studies have used the $BzK$-criterion to constrain the properties 
of star-forming and passive galaxies at $z \simeq 2$. For example, 
Kong et al. (2006) presented $K_s$-selected samples of $BzK$ galaxies over two fields: a $\simeq 920$ 
arcmin$^2$ field (with $K_{s; AB} < 20.8$) and a $\simeq 320$ arcmin$^2$ field (to  $K_{s; AB}\simeq 21.8$). 
They particularly concentrated their analysis on the clustering properties of $BzK$ galaxies, 
and concluded that the $pBzK$ galaxies are more clustered than the $sBzK$ galaxies. 
Meanwhile, Lane et al. (2007) combined the first 
release of the near-infrared UKIRT Infrared Deep Sky Survey (UKIDSS) 
Ultra-Deep Survey (UDS; Lawrence et al. 2007) with optical photometry from Subaru imaging. 
By comparing the commonly-used selection techniques 
for galaxies at intermediate redshift, they concluded that the brightest Distant Red Galaxies 
(DRG) have spectral energy distributions (SED) consistent with dusty star-forming 
galaxies at $z \simeq 2$. Moreover, they observed an interesting turn-over in their derived number counts 
of $pBzK$ galaxies at $K\simeq 21$, suggesting an absence of high-redshift passive galaxies 
at lower luminosities. Grazian et al. (2007), instead, focused on the
overlap between DRG and $BzK$ galaxies, discussing their relative
contribution to the overall stellar mass density.

The galaxy population in the UKIDSS UDS field was also studied by Hartley et al. (2008), 
in order to measure the clustering properties, number counts 
and the luminosity function of a sample of star-forming and quiescent $BzK$ galaxies with a 
limiting magnitude $K_{s; AB} <23$. The number counts they derived for the passive objects 
exhibit a flattening at  $K_s \simeq 21$ and an apparent turn-over at $K_s \simeq 22$:
they concluded that the former effect is likely to be real, 
while the latter is probable but remains uncertain. 
More recently, McCracken et al. (2010) presented 
number counts and clustering properties for a sample of $pBzK$ 
galaxies with  $K_s < 23$, selected over a significantly larger area than previous studies 
($2$\,deg$^2$ in the COSMOS 
field). They also found some evidence for a turn-over in the number counts 
of quiescent galaxies, around $K_s \simeq 22$.

It is interesting to notice that similar conclusions were already
  reached by other papers using different techniques.  For instance, De
  Lucia et al. (2007) used a combination of photometric and
  spectroscopic data to study the evolution of cluster environment  over a
  large range of cosmic time, from redshift 0.4 to 1. They found a
  significant deficit of low mass, faint red galaxies going to high
  redshift.  A similar conclusion, i.e. the truncation of the red
  sequence at faint magnitude for high redshift galaxies, was
  confirmed by other authors such as Kodama et al. (2004), and Andreon et
  al. (2011).  A decline in the number density of the quiescent population at faint magnitudes was  also observed in the $H_{160}$-band by Stutz et al. (2008). 
  More recently, Ilbert et al. (2013) presented galaxy
  stellar mass functions and stellar mass densities for star-forming and
  quiescent galaxies in the redshift range 0.2$<$z$ <$4 using the 
  UltraVISTA DR1 data release.  They studied the stellar mass function
  for both star forming and quiescent galaxies, 
  and found  that the number density of massive, quiescent galaxies
  is relatively unchanged out to high redshift. 
  They interpreted this
  evidence as a direct consequence of star formation being
  drastically reduced or quenched when a galaxy becomes more massive
  than M$>$ $10^{10.7-10.9}\,{\rm M_{\odot}}$.  This result is
  consistent with the one presented in Muzzin et al. (2013), who also
  computed the stellar mass function for quiescent galaxies using the
  UltraVISTA data.  In general, dedicated analyses of the evolution 
  of ``Red\&Dead'' galaxies show { an evolution in the fraction of red galaxies
   (e.g. Fontana et al. 2009, Brammer et al. 2011), that decreases as redshift increases}.

The use of the $BzK$ selection method for finding passively evolving galaxies has the distinct advantages of being relatively easy to perform on data, of being easily and 
self--consistently reproducible on mock catalogs derived from theoretical simulations, and has been relatively well tested with spectroscopic follow-up. However, it is crucial to stress that the success rate of the $BzK$ selection is strongly dependent on the availability of deep imaging at both optical and near-infrared wavelengths.  All the above-mentioned surveys were limited both by the depth of the near-infrared
data, and by the relative depth of the imaging in the bluer bands. The robust detection of passively-evolving $BzK$ galaxies, requires a proper measure of a color term as large as $B-K\simeq 7$.  If the $B$-band imaging is of limited depth, the clean selection of $pBzK$ galaxies cannot reach the faint $K$-band limits where the bulk of the population is expected to lie, and the observed number counts are severely prone to errors due to incompleteness and noise. Because of these limitations, the existence of a turn-over in the observed number counts of quiescent galaxies at $z \simeq 2$ has yet to be firmly established.

In this paper, we use a combination of wide-field and deep optical and infrared images
that allow us to robustly select a sample of passively-evolving galaxies at $z \simeq 2$ 
to $K_s \simeq 25$, and to explore the dependence of the number density
of these quiescent objects as a function of luminosity or, equivalently,
stellar mass.  The data have been acquired in the context of the
CANDELS HST (Grogin et. al 2011; Koekemoer et al. 2011) and HUGS VLT (Fontana
et al. 2014) surveys, that deliver a unique combination of
area and depth in the the $B$, $z$ and $K_s$ bands.
With this unique sample we study the number counts
of the $BzK$ galaxy population, and in particular we focus on the quiescent population
to provide improved constraints on 
the nature of the physical processes involved in the quenching of
star formation. 

The paper is organized as follows: in Section 2 we describe the data and the multiwavelength catalog for the UDS and GOODS-S fields, in Section 3 we provide an 
updated view of the properties of faint $BzK$ galaxies as obtained by the unique CANDELS data set, and in Section 4 we present the simulations 
performed to quantify the incompleteness of the observed sample. A reader less interested in the technical details 
may go directly to Sect. 5, where we present the crucial result of the paper, i.e. the luminosity and mass distribution of $pBzK$ galaxies, and to Sect. 6 , where we compare our results with the predictions of various semi-analytical models (SAM) of galaxy formation. Finally, a discussion of our results and a summary of our main conclusions is presented in Sec. 7.

All magnitudes utilised in the paper are in the AB system (Oke 1974) and a standard  cosmological model
($H_0$ = 70 km/s/Mpc, $\Omega_M=0.3$ and $\Omega_\Lambda=0.7$).

\section{Data}
\subsection{Imaging data}
\label{imagingdata}

The present paper is based on data collected over the CANDELS
pointings of the UDS and the GOODS-S fields.  Both these $\simeq 200$\,arcmin$^2$ fields have been observed in many broad bands in recent years, including WFC3/IR and ACS with the {\it Hubble Space Telescope} (HST), as a part of CANDELS HST Treasury Project (Grogin et al. 2011; Koekemoer et al. 2011). Additional ultradeep images from the $U$ to the mid-infrared Spitzer bands are also available. A complete description of the imaging data available in these fields along with the procedure adopted for catalog extraction are given in Galametz et al. (2013, G13 in the following) for UDS and in Guo et al. (2013, GUO13 in the following) for GOODS-S, respectively.

Since this work makes crucial use of the $B$, $z$ and $K$ bands to select passive galaxies, we briefly summarize here the properties of these images, 
including new imaging data in the $B$ and $K$ bands recently collected with VLT that are not included in the aforementioned catalog papers.

In the $K$ band, the two fields have been imaged by the High Acuity Wide field $K$-band Imager (HAWK-I) on the European Southern Observatory's Very Large Telescope (VLT) as part of the HUGS survey (an acronym for HAWK-I UDS and GOODS-S survey).  This program has delivered very deep images in the $K_s$ band over both fields, reaching 5-$\sigma$ detection limits fainter than $K_s \simeq 26$.  We refer to Fontana et al. (2014) for a complete description of the HUGS survey.  The $K_s$ band of HAWK-I will be referred to simply as the $K$-band in the rest of this paper.

For UDS, the final $K$-band photometry is already included in the G13 catalog. The whole UDS field is covered with three different pointings. Their seeing is $0.37 \prime \prime - 0.43 \prime \prime$  
and the corresponding limiting magnitudes are  $m_{lim}(K)\simeq 26$,  $m_{lim}(Y)\simeq 26.8$ (5$\sigma$ in one FWHM) or  $m_{lim}(K)\simeq 27.3$,  $m_{lim}(Y)\simeq 28.3$ (1$\sigma$ per arcsec$^2$). 
In GOODS-S, we use here the final HUGS image, that is considerably deeper than the image used in GUO13.
The whole field was covered with 6 (partly overlapping) pointings for a  total exposure time in the $K$ band (summed over the six pointings) of 107 hours. Because of the complex  
geometry, this corresponds to an exposure of 60--80 hours in the central area (the one covered by CANDELS ``Deep'' ) and 12--20  hours in the rest (the CANDELS ``Wide'' area). 
The final average seeing is remarkably good and constant, with 4 pointings at $0.38\prime \prime$ (notably including the two deepest) and 2 pointings at 
$0.42 \prime \prime$. On the finally stacked images, the limiting magnitudes in the deepest area are   $m_{lim}(K)\simeq 27.8$ (5$\sigma$ in one FWHM) or  $m_{lim}(K)\simeq 28.3$,  $m_{lim}(Y)\simeq 28.3$ 
(1$\sigma$ per arcsec$^2$).

The $B$ and $z'$ images in the UDS were obtained with Subaru/Suprime-Cam (see Furusawa et al. 2008). The Subaru optical data cover the full CANDELS UDS field, and the 5$\sigma$ 
limiting magnitudes (computed in 1 FWHM of radius) are $28.38$ and $26.67$ in $B$ and $z'$ respectively (G13).

Another crucial data set in this context is a recent ultradeep $B$-band image obtained with VIMOS on VLT for GOODS-S (Nonino et al. in prep.). Here
the average seeing is 0.89$\prime \prime$. With a VLT/VIMOS total integration time of 24\,hr, the final mosaics have a median PSF of $0.85$ arcsec FWHM, and 
reach $B = 28.4$ at 5$\sigma$, and are thus considerably deeper than the available ACS $F435W$ data (the latest V2.0 version released by the GOODS Team reaches a 5$\sigma$ limit of $F435W\simeq 26.9$). 
We will use in the following this VIMOS $B$-band image. As for the GOODS-S $z$ band, we used  the $F850LP$ image originally obtained in the GOODS program (Giavalisco et al. 2004) and then augmented by subsequent observations. Guo et al. 2013 report a $5\sigma$ limit of $m_{AB}=28.55$ in a small aperture of radius 0.09''. 

While the $B$, $z$ and $K$ bands are crucial for the selection of passive galaxies, we have also utilized the full multi-color data from the $U$-band 
to the mid-infrared Spitzer bands at various stages in our analysis, both to derive photometric redshifts and to determine the shapes of the spectral energy distributions (SEDs) 
of the selected galaxies. We refer to G13, GUO13, and Fontana et al. (2014) for a  complete description of the imaging data set.

\subsection{Catalogs}

The catalogs used in this paper are those already published in G13 and GUO13, with the only notable difference that we have included  the new $B$ and $K$ images, as described below.

The adoption of the CANDELS catalogs implies that we rely on the $H_{160}$ band for the detection of galaxies.  
This choice has the practical benefit that we have been able to use the existing CANDELS catalogs, and takes full advantage of the depth and the quality of the HST images. Because the $H_{160}$ imaging is still significantly deeper than the $K$-band imaging, this choice does not affect the definition of the $K$-selected sample; we have performed an independent detection in the $K$ band and verified that all the objects detected in $K$ are also detected in $H$-band, at least to the magnitude limits we are interested in here.

{ From the whole $H$--selected sample we have then selected the objects that fall in the regions where the $K$ band images are well exposed, trimming the outer edges of the images. The remaining areas are 151.08 arcmin$^2$ in GOODS-S (defined as all pixels of the image having a 1-$\sigma$ magnitude limit $>27$ in one sq. arcsec) and 178.09 arcmin$^2$ in UDS (made by pixels with a 1-$\sigma$ magnitude limit $>26.5$ in one sq. arcsec). The final UDS sample (that is complete to $H\simeq 26$ and extends to $H\simeq 27$) contains 27831 $H$--selected objects, while the GOODS sample  (that is complete to $H\simeq 26.6$ and extends to $H\simeq 28$) contains 31551 $H$--selected objects.}

Colors in the ACS and WFC3/IR bands have been measured by running SExtractor in dual-image mode, using isophotal magnitudes (MAG\_ISO) for all the galaxies, after smoothing each image with an appropriate kernel to reproduce the resolution of the $H_{160}$ WFC3/IR image. 
The Template-FITting 
photometry software TFIT (Laidler et al. 2007) was used to derive the photometry for all the images that have poorer resolution than $H_{160}$ (e.g.~$B$, $z'$ and $K$). 
This code uses information (position, profile) of sources measured on a high-resolution image (here $H_{160}$) as priors to establish the photometry in the lower-resolution images. We refer to G13 for a more detailed discussion of the CANDELS UDS multiwavelength catalog.
For the new $B$ and $K$ images used here we have applied TFIT exactly as described in G13 and GUO13.

We remark that the TFIT code that we used to measure colors is designed to minimize the effects of blending from nearby sources when extracting the photometry of faint sources in crowded extragalactic fields. The power of TFIT in this context is clearly shown by the simulations presented in Lee et al. (2012), where it is shown that the contamination from nearby sources is minimal even in the case of the $\simeq$3$\prime \prime$ 
PSF of the Spitzer images. The ground--based images on which this analysis is based ($B,z,K$) have a much smaller PSF, especially in $K$ where the seeing 
is consistently better than 0.45$\prime \prime$, and we therefore expect possible contaminations to be even smaller. As a further check, we have verified that only 5\% of galaxies at $K\simeq 23.5-24$ (the most typical objects that we target) have a nearby brighter companion within 2$\prime \prime$, that may be affecting the measured colors. We are therefore confident that our results are not significantly affected by photometric contamination.

We also note that the images used in this analysis are remarkably homogeneous in depth over the area that we use. Ground--based $B$ and $z$ images have been obtained with imagers with FOV larger than those of CANDELS fields, and the dithering pattern adopted for Hawk--I has ensured a proper coverage of the full area. We have explicitly verified this by looking at the distributions of the 1--$\sigma$ magnitude limits in the $B$, $z$ and $K$ 
images for our $BzK$ galaxies, and we verified that they are reasonably narrow gaussian distributions. The remaining inhomogeneities are accounted for in the simulations that we used to evaluate the systematics in our analysis.

\begin{figure}
  \centering
  \includegraphics[width=9cm]{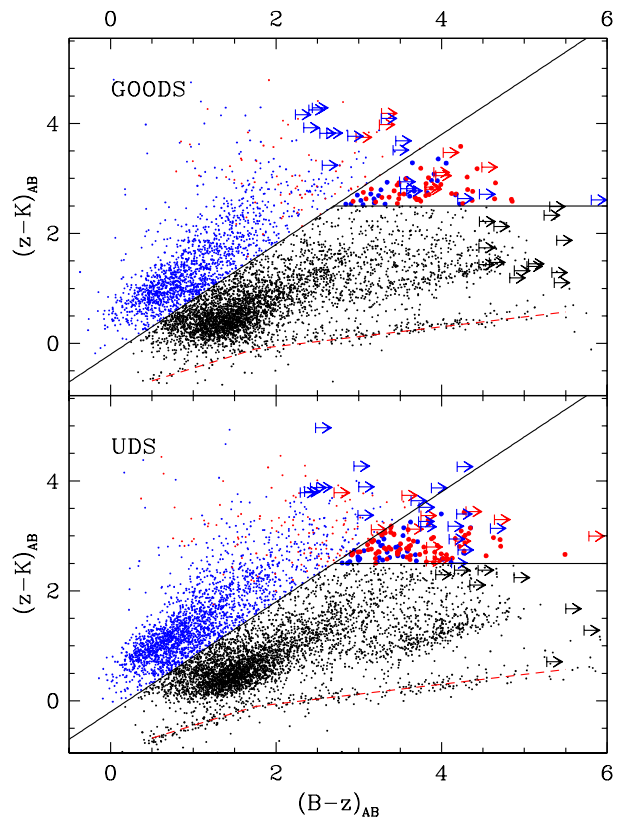}
   \caption{$B-z$ versus $z-K$ color-color diagram for the $K$-selected galaxies and stars in the 
   CANDELS GOODS-S (upper panel), and UDS (lower panel) fields. Arrows indicate $1\sigma$ 
   lower limits on the measured $(B-z)$ color for sources which are undetected (S/N $< 1$) in the $B$-band.
   Objects shown in blue are color-selected $BzK$ galaxies classified as ``star-forming'' from the SED analysis. Those plotted in red are color-selected $BzK$ galaxies classified as passively evolving with the same criterion. The red dashed line is a fixed reference to show the position of stars in the $BzK$ plane, useful to show the consistency of photometric calibration.
}
     \label{bzkuds}
  \end{figure}

To facilitate comparison of our results with 
previous studies we wanted our photometric selection criterion 
to match as closely as possible the original $BzK$ selection introduced in 
Daddi et al. (2004).
Since the filters used here in both the GOODS-S and UDS fields are not identical to those used in the
original $BzK$ paper, we computed the correction term to the colors to
account for the different shape of the filters.   Correction terms
were computed using synthetic models of Galactic stars, and checked against the terms computed with the Bruzual \& Charlot (2003) (BC03) evolutionary synthesis code.
Corrections for the two colors in GOODS-S and for $z-K$ in UDS turned out to be relatively small, nearly always less than 0.05 mag, both between the UDS and GOODS-S filters, 
as well as in comparison with the original Daddi et al. photometry. 
The $(B-z)$ color term in UDS turned out to be significantly larger, with a strong linear trend  up to $\Delta(B-z)_{UDS-GOODS-S}\simeq \Delta(B-z)_{UDS-Daddi}\simeq -0.3$. 
We therefore decided to apply only a correction to $(B-z)_{UDS}$, in order to bring it as close as possible to $(B-z)_{GOODS}$, neglecting all the other much smaller color terms.
The corrected color is $(B-z)_{UDS}^{corr}=(B-z)_{UDS} + 0.135\times (B-z)_{UDS}+0.045$ up to $(B-z)_{UDS}=2$, and $(B-z)_{UDS}^{corr}=(B-z)_{UDS}+0.315$ at larger $(B-z)_{UDS}$.
Hereafter we will apply this correction to data and theroretical models (when computed explicitly in the UDS filter set) and refer to $(B-z)_{UDS}^{corr}$ as $(B-z)_{UDS}$ for simplicity.
The final agreement can be visualized in Figure~\ref{bzkuds}, where we have plotted an identical red dashed line in the two panels that overlaps the stellar loci of the two fields. 
As can be seen, the consistency of the two calibrations is good.

\subsection{The SED fitting}



While the selection of passive galaxies is performed using only the $BzK$ bands, we exploit
the full multiwavelength catalogs delivered by CANDELS to obtain further information on the targeted galaxies, 
including photometric redshifts, stellar masses and other quantities.

For all the $BzK$ galaxies without a secure spectroscopic redshift we used the official photometric redshift 
produced by the CANDELS collaboration.  The technique adopted is described in Dahlen et al. (2013), 
and photometric redshifts for both fields will be made available in forthcoming papers (Dahlen et al., in prep.). 
For the sample of $BzK$ galaxies with secure spectroscopic redshifts we find that the scatter between photometric and spectroscopic redshifts is  gaussian distributed, with a standard deviation 
$\sigma\sim0.08$  and a small number of outliers,  about 5\%.

To compute the  physical parameters of the galaxies we used  a SED fitting technique applied
on the full multiwavelength  catalog available for the two CANDELS fields, that have 18/19 bands on UDS/GOODS, from $U$-band to 
8$\mu m$. 
The SED technique employed for this work  has been already intensively tested in previous papers 
(Fontana et al. 2003, 2004, 2006; Grazian et al. 2006; Santini et al. 2012). It
is based on the comparison between the observed multi-color SED of each 
galaxy with those obtained from a set of reference synthetic spectra from stellar population models.
The redshift of each galaxy is fixed to the spectroscopic or the photometric one during the fitting process.

We decided in this case to use a fairly standard reference spectral library, to facilitate 
comparison with results in the literature.
It is based on the  BC03 models, and the code was run using a Salpeter IMF, ranging over a set of metallicities
from $Z=0.02\,{\rm Z_{\odot}}$ to $Z = 2.5\,{\rm Z_{\odot}}$, and dust extinction $A_{V}$ 
ranging between 0 and 1.0 assuming a Calzetti
extinction  curve. We used  smooth exponentially-decreasing star-formation histories.
All the parameters adopted for the SED fitting are listed in  Tab.~\ref{sed}.
The best-fitting spectrum provides estimates of the star-formation rate (SFR), extinction, stellar population 
age (the onset of the star formation episode) and galaxy stellar mass,  and $\tau$ (the star-formation e-folding timescale).

Among the outputs of this analysis we will use in particular the stellar masses and the age/$\tau$ parameter, that is the inverse of the Scalo parameter.  As shown  in Grazian et al. (2007),  this parameter can be used to discriminate between star forming and passively evolving galaxies. Following their work, we assume age$/\tau<4$ for the former and age $/\tau\geq4$ for the latter.  

\begin{table}
\caption{Parameters used for the library of template SEDs.}       
\label{table2}   
\centering             
\begin{tabular}{c c}    
\hline\hline         
IMF 	& Salpeter \\  
\hline
SFR $\tau$ (Gyr) & 0.1,0.3,0.6,1,23,4,9,15\\
\hline
log (age) (yr) & 7, 7.01, 7.03...10.3\\
\hline
Metallicities & 0.02 Z$_{\odot}$, 0.2$_{\odot}$, 1$_{\odot}$, 2.5$_{\odot}$\\
\hline
E$_{B-V}$ & 0, 0.03, 0.06, 0.1, 0.15, 0.2, ..., 1.0\\
\hline
Extinction law & SMC, Calzetti\\
 \hline      
\end{tabular}
 \label{sed}
\end{table}

  \begin{figure}
  \centering
  \includegraphics[width=9cm]{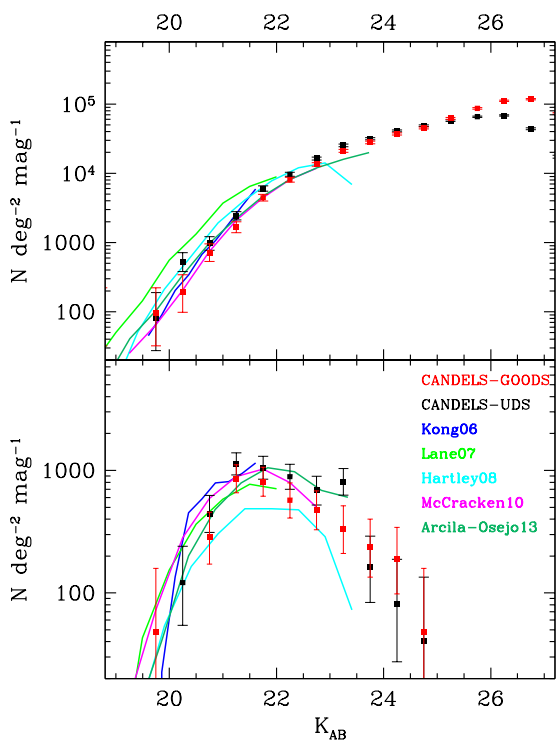}
   \caption{Number counts for the $sBzK$  (upper panel) and $pBzK$ (bottom panel) galaxies in the two CANDELS fields
   plotted separately, red GOODS-S and black UDS.
   We assume Poissonian
   error bars, computed with the Gehrels (1986)  formula for low number counts. Previous results from the literature are over-plotted:  Kong et al. (2006) in blue,
    Lane et al. (2007) in green, Hartley et al. (2008) in cyan, McCracken et al. (2010) in magenta,
    and Arcila-Osejo et al. (2013) in dark-green.
   }
  \label{RawCounts}
  \end{figure}


\section{$BzK$ galaxies in CANDELS} 
\subsection{Raw $BzK$ counts} 
As first proposed by Daddi et al. (2004), the $BzK$ color-color criterion is an efficient color-based 
method for identifying galaxies in the redshift range $1.4<z<2.5$. It also permits segregation of galaxies 
between actively star-forming galaxies ($sBzK$) and passively-evolving galaxies ($pBzK$). 
If  we define $BzK = (z-K) - (B-z)$,  $sBzK$ galaxies have colors consistent with $BzK> -0.2$ 
while $pBzK$ galaxies have $BzK<-0.2$ and $(z-K) > 2.5$ (corresponding to the upper-left 
and upper-right regions 
of the color-color diagram respectively; see Figure~\ref{bzkuds}).


Figure~\ref{bzkuds} shows the $B-z$ versus $z-K$ distribution of $K\leq 24$ sources in the CANDELS GOODS-S (upper panel) and UDS (lower panel) fields. While all $BzK$ galaxies are detected in the $z$-band images, the same does not hold for the B band, where several of  them are undetected even in our ultradeep images.   We assume for these sources a lower limit on $B$-band magnitude equal to their $1\sigma$ limiting magnitude, and plot them as horizontal arrows in  Figure~\ref{bzkuds}. We note that, in principle, $BzK$ galaxies undetected in the $B$ band and located in the $sBzK$ region could be $pBzK$ galaxies scattered left-ward into the $sBzK$ region, as shown also in Grazian et al. (2007), which therefore escape inclusion in the $pBzK$ sample. 

Based on our sample, we are able to compute the raw number counts derived for both the $sBzK$ and $pBzK$ galaxies in the CANDELS fields for both fields separately, as shown in Figure~\ref{RawCounts}. The error bars are computed assuming  Poisson statistics,
approximated with the square root of the counts  for $N>10$ and the small-number approximation 
for the Poisson distribution of Gehrels (1986) for $N\leq10$.  It is immediately clear that our observations reach a depth that no previous studies have achieved.  Also evident is the 
decrease in the number counts of $pBzK$ galaxies at magnitudes fainter than $\simeq 21.5-22$. To establish whether this turnover is real we have performed dedicated simulations that are described in the following section.

The number counts for star-forming and passive  $BzK$ galaxies are reported in Table~\ref{ncSF}
and Table~\ref{nc}, respectively.  Since the detection is made on the deep $H$-band CANDELS images,  the number counts for the $sBzK$ extends well below $K\simeq 27$, where they are however significantly plagued by incompleteness in the detection. Since we are interested here in the $pBzK$ population, that is limited at much brighter magnitudes by incompleteness due to color selection, as we shall quantify below, we have made no attempt to correct for the detection incompleteness in the faintests bins of the $sBzK$ counts.

As far as $sBzK$ galaxies are concerned, our number counts are in reasonable agreement with most of the previous studies, encompassing the range of densities found by previous works, that typically were conducted over much larger areas. Probably the most striking difference is with respect to Lane et al. (2007), whose counts are above our own (and all the others).  As proposed by McCracken et al. (2010), this difference could be due to an inappropriate transformation to the Daddi et al. filter set. However, we also note that they have been obtained over the UDS region. As shown in Figure~\ref{RawCounts}, our counts in the UDS region are also above those in the GOODS-S region.  It is therefore possible that some of this excess is due to a true overdensity of galaxies in the UDS field, where indeed a large number of $z\simeq 1.6$ galaxies 
within a high-density structure has been observed (Papovich et al. 2010; Tanaka et al. 2010). We also note that the mismatch between our two field is close to the limit of 
what can be accounted for by a straightforward application of clustering statistics. Looking for instance at $sBzK$ galaxies with $K\leq 22$, we find $151$ and $248$ galaxies in the GOODS-S and UDS fields, respectively. Using the Cosmic Variance Calculator (Trenti \& Stiavelli 2008) to estimate their expected variance and converting their density to the same area of one square degree, we find that their densities are $ 3600 \pm 596$ galaxies per square degree and $5015 \pm 707$ galaxies per square degree, repectively, hence discrepant at the $2-3\sigma$ level.

Focusing our attention on the main subject of the current paper, 
the  number densities of $pBzK$ galaxies in the two fields match the observed number densities from the literature in the bins where we have good statistics. 
There is some discrepancy with
previous works, especially with McCracken et al. (2010) and Kong et al. (2006). This difference  may  be due to cosmic variance, given the strong clustering of p$BzK$
galaxies with respect to $sBzK$ galaxies. There is also a systematic difference between our two fields:  in total, we identify 134 $pBzK$ galaxies in the UDS and 81 in GOODS-S: even in this case, the UDS field seems overdense. Following the  Cosmic Variance Calculator as described before, this translates into a number density of $1930\pm380$ and $2710\pm440$ in the GOODS-S and UDS field, respectively, that is still consistent within $2\sigma$.



\subsection{The redshift distribution of $BzK$ galaxies}
\label{sec:red}
\begin{figure}
  \centering
  \includegraphics[width=9cm]{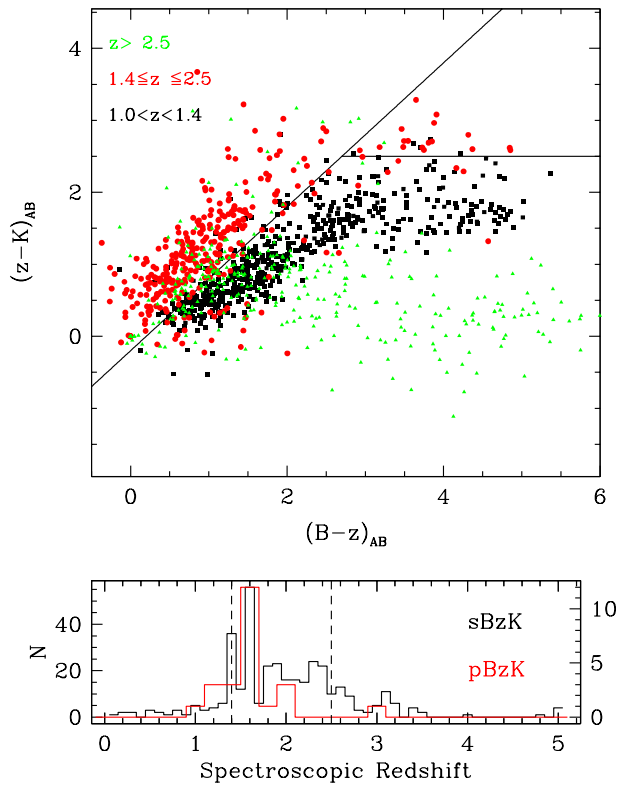}
   \caption{$BzK$ with spectroscopic
   redshift. {\it Upper panel: } Position of all galaxies with $z_{spe}>1$ on the  $BzK$ plane. Each symbol refers to a different 
   redshift range, as shown in the legend.  {\it Lower panel: } 
   The spectroscopic redshift distribution for the $BzK$ galaxies, divided between $sBzK$ and $pBzK$, as shown in the legend. The vertical dashed lines represent the canonical redshift range of $BzK$ galaxies.}
     \label{zspe}
  \end{figure}

Both the UDS and the GOODS-S fields have been the target of extensive spectroscopic surveys in the past. Using these publicly available data we have first verified the accuracy and completeness of the $BzK$ selection criteria in selecting galaxies in the redshift range $1.4<z<2.5$, expoiting a sample that is significantly larger than the original K20 sample used by Daddi et al. (2004).  This is 
shown in Figure~\ref{zspe}, where we plot the position of { all} sources with spectroscopic redshift $z_{spe}>1$ in the $BzK$ diagram; each symbol refers to a different redshift range. Figure~\ref{zspe} also shows the distribution of the spectroscopic redshifts of these $BzK$--selected galaxies. The significant overlap between the $1.4<z<2.5$ galaxy population and the $BzK$ selection criterion is immediately evident.  More quantitatively, we find that
within the spectroscopic sample, 318 galaxies (38 \& 280 in the UDS and GOODS-S, respectively) fall in the $sBzK$ region. Of these, 225 (23 \& 202) are at $1.4\leq z \leq 2.5$). The total number of $pBzK$ galaxies
with spectroscopic redshifts is 15 (1 \& 14), with only two of them outside of the $1.4\leq z \leq 2.5$ redshift range. Conversely, 56 galaxies (2 \& 54) at $1.4\leq z \leq 2.5$ are not in the overall  
$BzK$ region. Based on these numbers we conclude that the completeness of the $BzK$ selection criterion is about 80\%, and the purity around 70\%. We have also verified that these numbers drop 
very significantly as soon as we move below $z=1.4$ or above $z=2.5$. This comparison confirms that the $BzK$ criterion is a very effective method to select galaxies in the specific redshift range $1.4<z<2.5$ 

\begin{figure}
  \centering
  \includegraphics[width=9cm]{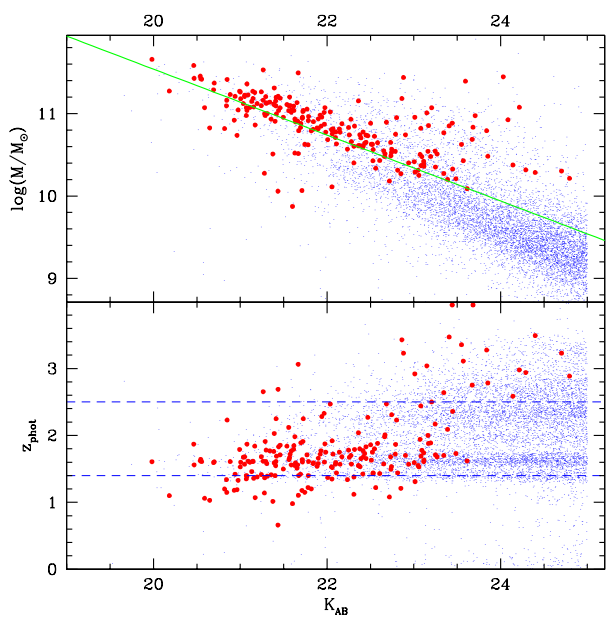}
   \caption{Distribution of  photometric redshift and stellar masses (upper panel) as a function of the $K$ magnitude for the whole $BzK$ sample. Blue small points refer to $sBzK$ galaxies, red dots to $pBzK$. In the lower panel the two dashed lines represent the nominal redshift range of the $BzK$ distribution; in the upper panel, the green thick line represents the average $M_{\odot}-K$ relation of Daddi et al. (2004). }
     \label{hisph}
  \end{figure}

{ Despite the extensive spectroscopic coverage, for 95\% of our $BzK$ sample we must rely on the accurate photometric redshifts to estimate rest-frame properties like stellar mass. 
The resulting picture is shown in Figure~\ref{hisph}, where we plot both the photometric redshift as well as the stellar mass of all $BzK$ galaxies as a function of their $K$ magnitude (the two fields are shown together). 
It is easily appreciated that most of the $pBzK$ galaxies have
photometric redshifts below $z=2$, with a  $z_{med}\simeq 1.8$,
while the distribution of $sBzK$ spans a wider range with $z_{med}\simeq 1.9$. 
The redshift distribution of the passively-evolved galaxies  is in good 
agreement with the one presented in McCracken et al. (2010).
We note that the peak of galaxies around $z\sim1.6$ could be related to  
large-scale structures known in both UDS (Papovich et al. 2010; Tanaka et al. 2010) 
and GOODS-S fields (Castellano et al. 2007; Kurk et al. 2008).

The central point of Figure~\ref{hisph}, however, is the evidence that the majority of the faint $pBzK$ galaxies - those at $K>23$ -  are located at $z_{phot}>2.5$, at variance with the brighter $K<23$ sample, that are as expected at $1.4<z<2.5$. We have individually inspected these sources and found that their photometry is not corrupted, and their photometric redshifts are reasonably well constrained. We note that the presence of passively evolving galaxies in the $pBzK$ region is not in contradiction with the expectations of synthesis models: the BC03 spectral library described before predicts that also passively evolving galaxies at $z>2.5$ may also be located in the $pBzK$ area. 

It is also interesting to look at the stellar mass  estimates for the $BzK$ population. 
As shown in the upper panel of Figure~\ref{hisph}, both $pBzK$ and $sBzK$ follow a general trend between the $K$-band and the stellar mass $M_{\odot}$. Indeed, Daddi et al. (2004) derived an average relation between $K$-band magnitude and
the stellar mass, of the form:
$log(M_{\star}/10^{11}\,{\rm M_{\odot}}) = -0.4(K^{tot} - K^{11}), $
where $K^{11}= 21.35$ (AB) is the $K$-band magnitude corresponding on average to a mass of $10^{11}\,{\rm M_{\odot}}$.
This was calibrated on the stellar mass estimates derived from full SED fitting in the K20 spectroscopic sample (Fontana et al. 2004), i.e. on galaxies with $K_{AB}$-magnitude less than 22, where exactly the same numerical code adopted here was used to estimate masses, although a previous version of the BC03 code was adopted.
This is shown as a green thick line in Figure~\ref{hisph}. 
The superior quality and spectral extension of the CANDELS imaging data gives us the opportunity of revising 
such relation.
While the Daddi et al. (2004) relation seems to fairly represent the correlation for bright sources, we find that it systematically overpredicts the stellar mass for fainter sources; 
this is consistent with expectation considering that faint $sBzK$ are typically low-mass star-forming galaxies, whose $M/L$ is expected to be lower than more massive, evolved galaxies.

It is interesting to note that the faint $pBzK$ galaxies that we locate at $z>2.5$ are significantly offset from the general relation, as expected since they are at considerably larger redshifts. }

  \begin{table}
\caption{ Observed Number Counts of $sBzK$ and $pBzK$ galaxies in the UDS and GOODS-S fields, separately. The total averaged number counts, including also a completeness correction, are given in Table 3. The first column shows the center of each 0.5~mag bin.}      

\centering             
\begin{tabular}{c c c c c c}    
\hline\hline         
$K_{AB }$ & $N_{ sBzK-UDS}$ & $N_{ sBzK-GOODS}$ & $N_{ pBzK-UDS}$ & $N_{ pBzK-GOODS}$\\  
\hline\hline           
 
   17.75 & 0 & 1 & 0 & 0 \\ 
  18.25 & 0 & 1 & 0 & 0 \\ 
  18.75 & 0 & 0 & 0 & 0 \\ 
  19.25 & 0 & 0 & 0 & 0 \\ 
  19.75 & 2 & 2 & 0 & 1 \\ 
  20.25 & 13 & 4 & 3 & 0 \\ 
  20.75 & 24 & 15 & 11 & 6 \\ 
  21.25 & 60 & 35 & 28 & 18 \\ 
  21.75 & 149 & 93 & 26 & 17 \\ 
  22.25 & 239 & 169 & 22 & 12 \\ 
  22.75 & 404 & 282 & 17 & 10 \\ 
  23.25 & 625 & 441 & 20 & 7 \\ 
  23.75 & 778 & 586 & 4 & 5 \\ 
  24.25 & 1012 & 776 & 2 & 4 \\ 
  24.75 & 1207 & 945 & 1 & 1 \\ 
  25.25 & 1421 & 1302 & 0 & 0 \\ 
  25.75 & 1623 & 1816 & 0 & 0 \\ 
  26.25 & 1690 & 2342 & 0 & 0 \\ 
  26.75 & 1089 & 2471 & 0 & 0 \\ 

 \hline      
\end{tabular}
 \label{ncSF}
\end{table}

%


\section{Estimating the systematic effects}
\label{valsim}

\subsection{Incompleteness}
\label{incom} 
As is evident by looking at Figure~\ref{RawCounts}, the number counts of $pBzK$ galaxies
 drop at fainter magnitudes beyond $K\simeq 21.5$. Since this is the main focus of the present paper, 
 before analyzing it in more detail it is important to exclude the possibility that this behavior is due to systematics in the data.  
As already discussed in Sect. 2, we can exclude that blending is a major cause of photometric errors, or that image inhomogeneities are a major source of incompleteness. 
It is certainly possible, however, that  this effect is due to photometric scatter. 
{ The photometric scatter could be potentially troublesome in the case of the $B$ band, which ultimately limits our $pBzK$ 
selection at the faint end. To properly measure the $B-K$ color of a  galaxy at the center of the $pBzK$ region (e.g. with $(B-z)=4$ and $(z-K)=3$)  
the $B$ band must be deep enough to allow the detection of the object as much as $\simeq 7$ magnitudes fainter than the $K$-band magnitude 
of the object. Given the depth of our $B$-band ($B\simeq 30$\,mag.}) we expect that incompleteness is already important at $K\simeq 23$. Previous analyses were done using 
$B$-band images about 2 magnitudes shallower, and we can therefore expect these effects to have been in place at even brighter magnitudes in those surveys.   
To further verify this, we performed  simulations that are described below.

\begin{figure}
   \centering
   \includegraphics[width=9cm]{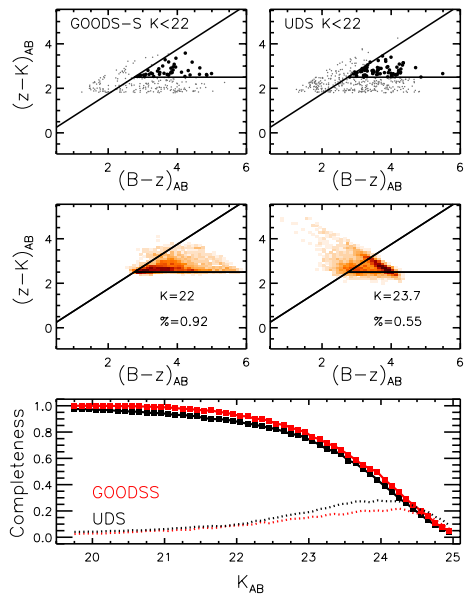}

      \caption{Simulations used to estimate the completeness in the selection of $pBzK$ galaxies. {\it Upper panel:} Position of the sources used to estimate the incompleteness in the two fields, taken as real $pBzK$ galaxies at $K<22$. {\it Central panels:} Example of the density of simulated sources in the $BzK$ plane, when sources are scaled to $K=22$ (left) and $K=23.7$ (right), and the effect of noise is included (see text for details). The case of GOODS-S is shown in this example. {\it Lower panel:} Big dots delineate the completeness (i.e. the  fraction of objects originally located in the $pBzK$ region that  are recovered even when the appropriate photometric noise is included)} as a function of the $K$-band  magnitude resulting from the full simulations, separately for the two fields. The small dots show the contamination, i.e. the fraction of objects detected in the $pBzK$ region that were originally located outside the same region.
         \label{simobs}
   \end{figure}

{ The effects of noise and incompleteness can impact in various ways the observed distribution of galaxies in the $BzK$ plot. First, photometric noise may scatter galaxies within the $BzK$ diagram: in particular galaxies that lie close to the lines that define the selection criteria can migrate in or out of the selection windows.  In the case of the horizontal line at $z-K>2.5$ the exquisite depth of the CANDELS and HUGS data is crucial to keep these effects to a minimum. For instance, at $K=24$ in the GOODS-S field the typical error on the $K$-band magnitude is $\simeq 0.05$ mag, and the color term $z-K=2.5$ is measured with a total error of typically 0.2 mag (we 
emphasize that all the p$BzK$ galaxies are detected in the $z$-band).  The diagonal selection line is more affected by the depth of the $B$-band image. Again, the depth of the $B$ band (the $1\sigma$ detection limit is $B\simeq 30$ in both fields) ensures detection also in the $B$-band for a large fraction of the $pBzK$ galaxies down to $K=24$. Those undetected are shown as arrows in Figure\ref{bzkuds}: those lying outside 
the $pBzK$ region could be in principle bona-fide $pBzK$ galaxies scattered out because of the inadequate depth of the $B$-band.

Clearly, the opposite effect may also happen: galaxies located outside the boundaries  of the $pBzK$ criterion can enter the region because of photometric noise and be included in the $pBzK$ galaxy sample.
In order to quantify these effects in our samples, we performed a
dedicated set of simulations. The ultimate goal is to define a robust
limiting magnitude in the $K$-band, and to correct
a-posteriori the observed number counts for  {\it incompleteness} (i.e. the fraction of true $pBzK$ galaxies lost due to noise) and  {\it contamination} (i.e. the fraction of galaxies selected by the  $pBzK$ criterion that were instead originally located outside it) .

The procedure that we followed is described in  Figure~\ref{simobs}. To estimate the incompleteness, we  started  from the sample of  observed $pBzK$ galaxies, using the brightest part of the catalog, 
where observational errors are so small that they do not contribute significantly to the observed scatter. After inspection of the error budget in the input catalogues, we decided to use for this purpose all objects with 
$K<22$. 
Similarily, we use the observed catalog of objects with $K<22$ and located in a region around the $pBzK$ boundary to estimate the amount of contamination in our sample.
This sample is shown in the upper panel of Figure~\ref{simobs}, where we plot the distribution of the objects used as input in the $BzK$ plane, that is slightly different in the two fields. We show both the galaxies selected as input $pBzK$ galaxies (big dots) as well as those lying immediately out of the $pBzK$ region (small dots) used to estimate the contamination.

We use these objects 
to generate mock catalogs of $10^{5}$ objects for each 
field, normalized at various $K$-band magnitudes from $K=20$ to $K=25.5$.
The computed magnitudes (in all bands) are then perturbed assigning
them a noise consistent with the observed error-magnitude relation in
each band and field. 
Our simulations accurately reproduce not only the average S/N as a function 
of magnitude, but also its scatter in order to match as closely as possible 
the observed and simulated data (Castellano et al. 2012).

Both these  catalogs are then analyzed in a similar way to the ones derived from the observations.
The output of this exercise is shown in the central panel of Figure~\ref{simobs}, where we plot the position of the simulated catalogs after applying the noise, for $pBzK$ galaxies only, in the case of two reference magnitudes ($K=22$ and $K=23.7$). For simplicity we plot only the GOODS-S field. 

This plot clearly shows the effect of the finite depth of our observations. As galaxies become fainter, there is some scatter across the $z-K=2.5$ boundary, that is however relatively small. By far the larger effect is the migration of objects toward smaller  $B-z$, onto the clear locus of galaxies with undetected B-band flux. This effect is clearly dependent on the object flux and removes a significant fraction of objects from the $pBzK$ region. Similar results (not shown in the figure) are found regarding the contamination - here the primary effect is due to photometric noise that causes objects close to the boundaries to enter the $pBzK$ region.

The final results are shown in the lower panel of the same Figure~\ref{simobs}, where we plot both the completeness and the contamination as a function of the $K$-band magnitude. They have been computed separately for the two fields, that have slightly different depths in the various bands.
 We find that the
incompleteness obviously increases with increasing magnitude, being relatively small ($<20$\%) up to $K=22.5-23$ but reaching a level of 50\% at about $K=23.5$ for UDS and $K=24$ for GOODS-S. 
In the following we will therefore focus our analysis on the $K<24$ sample, and simply present our results down to $K=25$, warning the reader that incompleteness corrections are very severe at $K>24$. The contamination also increases at the faint end of our sample, reaching 20-25\% at $K\simeq 24$. We will use these curves in the following to correct our number counts. }

\subsection{Spectral classification}
\label{sedfit}
We have also compared the galaxies selected with the simple color 
selection criteria to the output of a full SED analysis, performed on
the multi-wavelength catalogs obtained from CANDELS. 
To understand the limitations of the $BzK$ selection  criterion,
derived only from  observed colors,
we follow a different approach based on the spectral fitting technique.
One simple test to verify the validity of the $BzK$ criterion is to analyze the population of $BzK$ galaxies by correlating the age $t$, the star-formation 
timescale $\tau$ and the dust extinction $E(B-V)$ derived from the SED fitting technique already described in Sec. 2.3.  

As in Grazian et al. (2007),  for each galaxy we compute the  $t/\tau$
parameter, and we characterize as star-forming the galaxies with $t/\tau<4$, and 
as passively-evolving those with $t/\tau\geq4$.  

Therefore, we first verify 
that the classification based on the $t/\tau$  parameter is consistent
with the results obtained using the $BzK$ criteria, i.e. we compute this parameter for
all the $BzK$ galaxies,  
 and for all of them  we examine the correlation 
 between  $t/\tau$  and the extinction. We verify that both star-forming and 
 quiescent galaxies have the expected value of $t/\tau$, as well as for the extinction. 
As shown in Fig.~\ref{bzkuds}, there is a general consistency between the position in the $BzK$ plane and the result of the SED fitting (we remind that the latter is obtained over 18-19 bands, not only on 3 as in the $BzK$ case).
Those located in the
$pBzK$ region are mostly fitted  with passively evolving models,
while those located in the $sBzK$ area are fitted with star--forming
models. This supports our conclusion that a relatively low number of
$pBzK$ galaxies are scattered out of the $pBzK$ region because of
limited depth of the $B$-band imaging.

\section{Passively evolving BzK galaxies: luminosity and mass distribution}
\label{mainres}

  \begin{figure*}
  \centering
  \includegraphics[width=9cm]{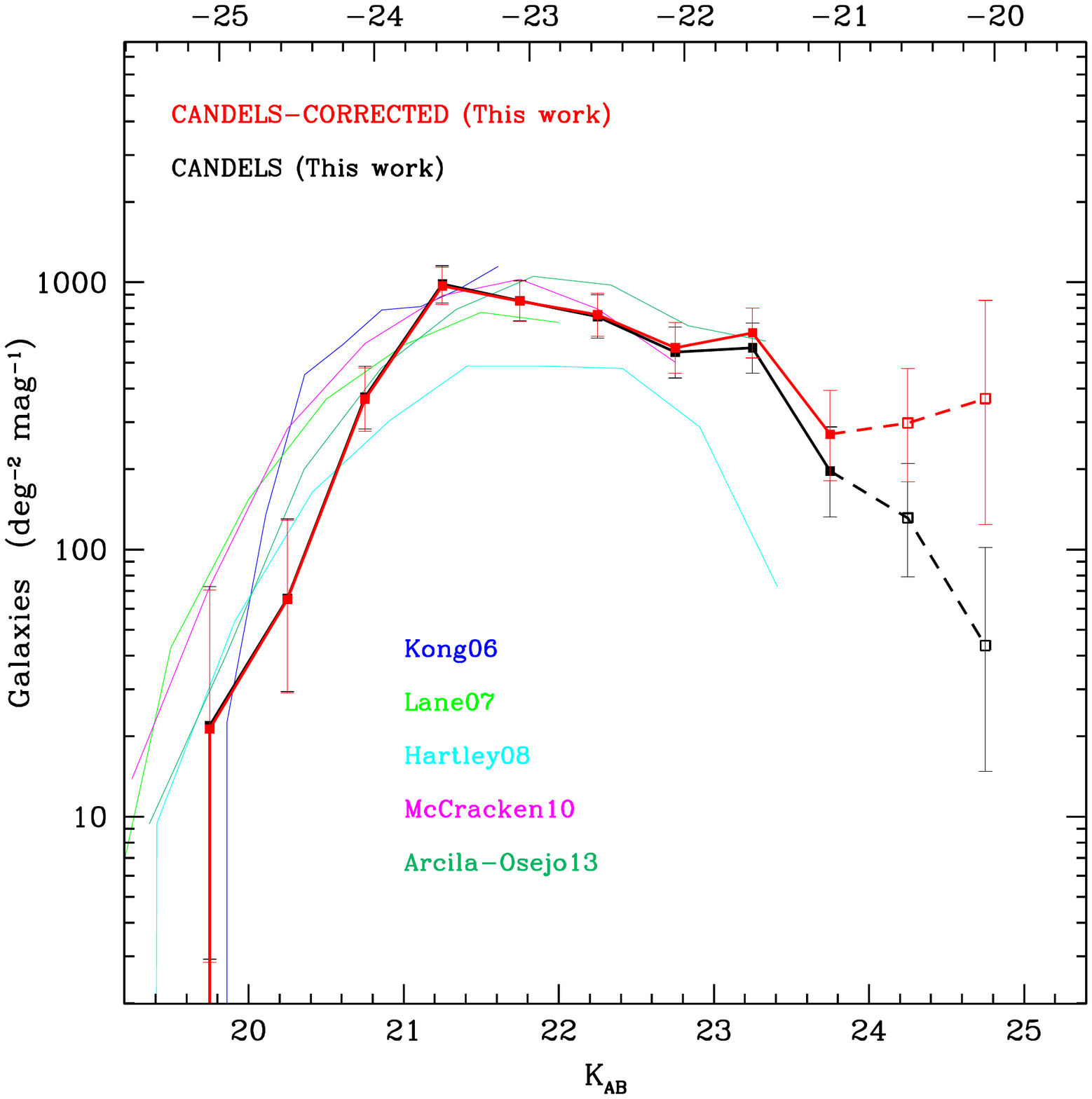}
  \includegraphics[width=9cm]{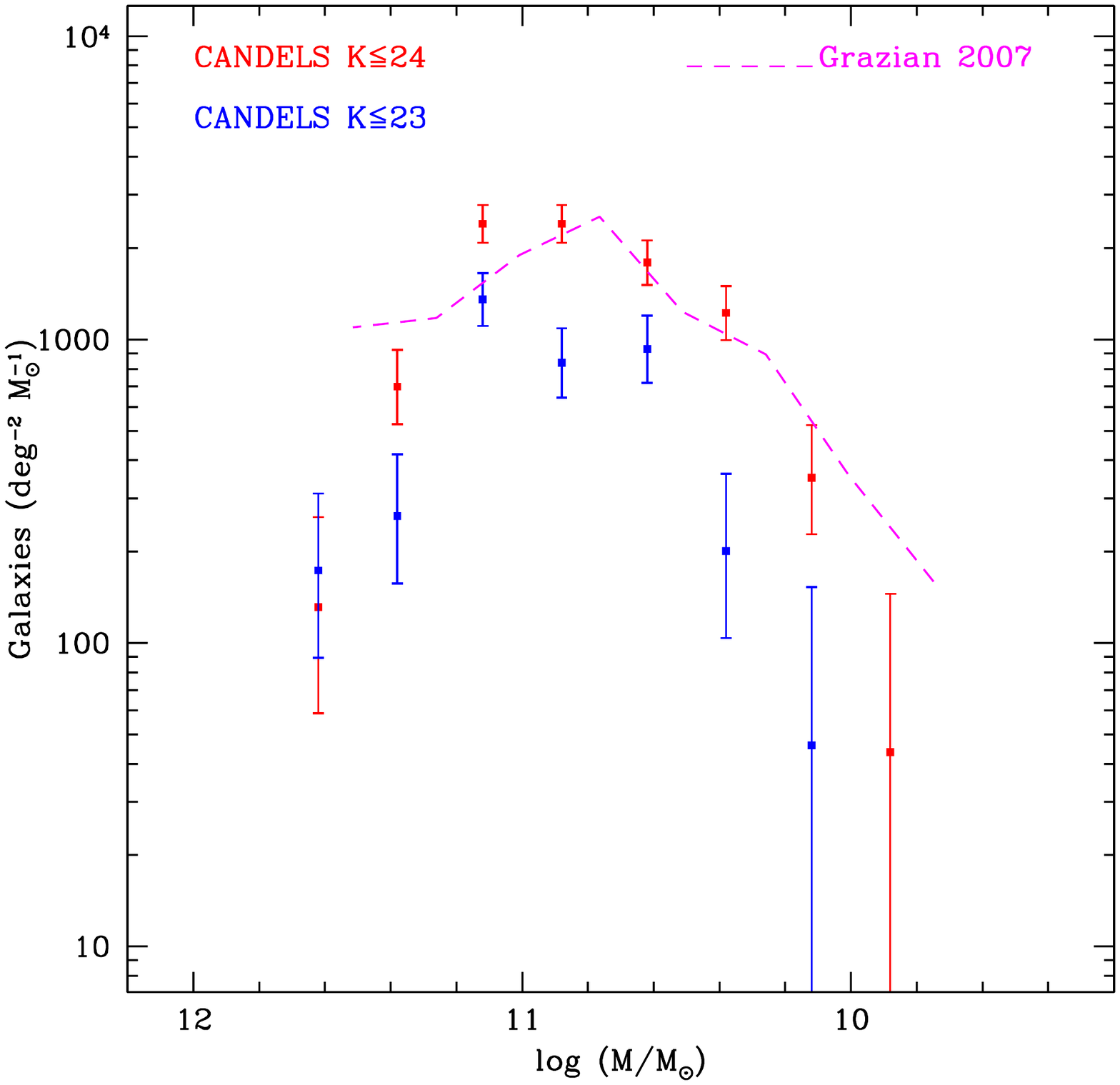}
     \caption{{\it Left}: The number counts of the CANDELS (UDS+GOODS-S) $pBzK$ galaxies
       as a
       function of the observed K magnitude. The black points and line show the raw data, 
while the red points and line show the data after correcting for incompleteness as discussed in the text. 
The upper x-axis shows the
       corresponding rest-frame absolute magnitude in the $I$ band at
       $z\simeq 2$. { As highlighted in the text,
       the corrections for incompleteness are below a factor 2 at most up to $K\simeq 24$, that sets the limit for a rigorous analysis. 
	 However we plot the faint end of the distribution until $K<25$ with a dotted line for reference.}
        The 	blue, green, cyan, magenta and darkgreen lines show the number counts for $pBzK$ galaxies from the literature, from
	Kong et al. (2006), Lane et al. (2007), Hartley et al. (2008),  McCracken et al. (2010), and  Arcila-Osejo et al. (2013) respectively. 
	{\it Right}: The stellar mass distribution of the CANDELS $pBzK$ galaxies, 
        after correction for incompleteness and contamination as described in the text. Masses are obtained with the multi-band SED-fitting technique.  
	Blue points are obtained from the $K<23$ sample, where all galaxies are at $z<2.5$, while the red points are from the $K<24$ sample, that includes also objects up to $z\simeq 3.5$. For comparison we plot also
	the galaxy stellar mass distribution of the $pBzK$ derived by Grazian et al. (2007) from the GOODS-MUSIC 
        sample (magenta dashed line).
      }
     \label{Pcounts}
  \end{figure*}

Armed with full characterization of the selection function,
we are now in a position to explore the properties of the $pBzK$ galaxies.
Figure~\ref{Pcounts} presents the $K$-band number counts of the $pBzK$ galaxies in the 
CANDELS fields. In addition to the raw counts, averaged over the two fields, we present the 
number counts corrected for incompleteness, up to $K=24$. For this purpose each galaxy of magnitude $K$ in the sample has been summed 
as $[1-contamin(K)]/complet(K)$, where $contamin(K)$ and $complet(K)$ are the contamination and completeness
as defined above and shown in the lower panel of Figure~\ref{simobs}. Error bars are Poissonian and computed as described above.
For comparison, we also overplot the number counts found by Kong et al. (2006),
 Lane et al. (2007), Hartley et al. (2008), and McCracken et al. (2010), that were not corrected for incompleteness.
The number counts are in good agreement with previous studies at faint magnitude. 

The $pBzK$ galaxy number counts derived from our CANDELS (UDS+GOODS-S) 
sample are summarized in Table~\ref{nc}.

When the correction is applied, the observed number counts still show a flattening at $K > 21$, consistent
with past studies (e.g. Hartley et al. 2008). The turn-over at $K \simeq 22$ is still present, although somewhat less significant. 
Converting these magnitudes to average rest-frame magnitudes in the $I$ band, (which is sampled by the $K$ band at $z\simeq 1.9$, close to the average redshift of $pBzK$ galaxies with spectroscopic redshift ) this corresponds to absolute magnitudes of $M_I \simeq -23$ and $M_I \simeq -22$, respectively, as shown by the upper x-axis in Figure~\ref{Pcounts} .

To further investigate the physical significance of 
this trend, we translate the galaxy photometry 
into stellar mass. 
We compute the stellar masses using our SED fitting to the
full multi-wavelength catalogs available for the two fields, { computed as described
in  Sections 2.3 and 4.2. We remind the reader that we adopt here a Salpeter IMF}.
 
 

\begin{table}
\caption{ Differential Number Counts for $sBzK$ and  $pBzK$ averaged over the (UDS$+$GOODS-S) fields. 
The first column is the center of the bins. The remaining columns are  in units of log(N/deg$^{2}$/mag) . 
The last column shows the $pBzK$ counts after correction for incompleteness and contamination, as described in the text.}        
\label{counts_tot}   
\centering             
\begin{tabular}{c c c c }    
\hline\hline         
$K_{AB }$ & $N_{ sBzK}$  &$N_{ pBzK}$ &  $N_{ pBzK}^{Corr}$ \\  
\hline\hline           

 19.75 &    1.817$^{   0.30 } _{   0.35 } $ &    1.340$^{   0.52 } _{   0.87 } $ &    1.327$^{   0.52 } _{   0.87 } $ \\ 
 20.25 &    2.570$^{   0.12 } _{   0.12 } $ &    1.817$^{   0.30 } _{   0.35 } $ &    1.813$^{   0.30 } _{   0.35 } $ \\ 
 20.75 &    2.920$^{   0.08 } _{   0.08 } $ &    2.570$^{   0.12 } _{   0.12 } $ &    2.563$^{   0.12 } _{   0.12 } $ \\ 
 21.25 &    3.304$^{   0.05 } _{   0.05 } $ &    2.993$^{   0.07 } _{   0.07 } $ &    2.987$^{   0.07 } _{   0.07 } $ \\ 
 21.75 &    3.696$^{   0.03 } _{   0.03 } $ &    2.931$^{   0.07 } _{   0.08 } $ &    2.930$^{   0.07 } _{   0.08 } $ \\ 
 22.25 &    3.935$^{   0.02 } _{   0.02 } $ &    2.871$^{   0.08 } _{   0.08 } $ &    2.879$^{   0.08 } _{   0.08 } $ \\ 
 22.75 &    4.165$^{   0.02 } _{   0.02 } $ &    2.738$^{   0.09 } _{   0.10 } $ &    2.755$^{   0.09 } _{   0.10 } $ \\ 
 23.25 &    4.354$^{   0.01 } _{   0.01 } $ &    2.755$^{   0.09 } _{   0.09 } $ &    2.811$^{   0.09 } _{   0.09 } $ \\ 
 23.75 &    4.461$^{   0.01 } _{   0.01 } $ &    2.294$^{   0.16 } _{   0.17 } $ &    2.431$^{   0.16 } _{   0.17 } $ \\ 
 24.25 &    4.579$^{   0.01 } _{   0.01 } $ &    2.118$^{   0.20 } _{   0.22 } $ &    2.473$^{   0.20 } _{   0.22 } $ \\ 
 24.75 &    4.661$^{   0.01 } _{   0.01 } $ &    1.641$^{   0.37 } _{   0.47 } $ &    2.565$^{   0.37 } _{   0.47 } $ \\ 
 25.25 &    4.760$^{   0.01 } _{   0.01 } $ &  ~  &  ~  \\ 
 25.75 &    4.860$^{   0.01 } _{   0.01 } $ &  ~  &  ~  \\ 
 26.25 &    4.924$^{   0.01 } _{   0.01 } $ &  ~  &  ~  \\ 
 26.75 &    4.873$^{   0.01 } _{   0.01 } $ &  ~  &  ~  \\ 

 \hline      
\end{tabular}
 \label{nc}
\end{table}

Figure~\ref{Pcounts} shows the resulting distribution of the stellar masses for the $pBzK$ galaxy
  sample computed with the SED fitting technique.  As shown in Figure~\ref{hisph}, the majority of $pBzK$ galaxies at $K>23$ appear to lie at $z>2.5$, at variance with those at $K<23$. We therefore plot separately the distribution at $K<23$ and at $K<24$.
 
  For comparison, we overplot the results obtained by Grazian et al. (2007), that were obtained from the $pBzK$ sample in the (less deep) GOODS-MUSIC sample using the same SED fitting technique.  The two results are quite consistent.

The observed distribution shows that there is a clear decrease in the
number density of passively
  evolving galaxies at stellar 
  masses below $10^{10.8}\,{\rm M_{\odot}}$.  
  This trend departs significantly from the overall form of the galaxy stellar mass function at this redshift, which continues to rise steeply to much lower masses (e.g. Ilbert et al. 2013).  We stress that this implies that the fraction of $pBzK$ galaxies compared to all $BzK$ galaxies, and hence to most of the galaxies at $z\simeq 2$,  is large ($\sim$25\%) at $M_* \simeq 10^{11}\,{\rm M_{\odot}}$ and becomes minimal ($\sim$1\%) at $M_* \simeq 10^{10}\,{\rm M_{\odot}}$.

\section{Comparison with semi analytical models of galaxy formation}
In the previous sections we have highlighted the existence of a clear
break in the luminosity and mass distribution of passive galaxies at
$z\simeq 2$, with these objects becoming progressively rarer beyond 
an observed $K$-band magnitude of $K \simeq 22$. This turnover in the 
number counts can be translated into a mass threshold roughly placed around
$10^{10.8}\,{\rm M_{\odot}}$. This implies that the mechanisms that halted star
formation in galaxies at high redshift have been more
efficient (or statistically more frequent) in massive galaxies rather
than in lower mass ones. 
It is interesting to investigate whether this basic feature is 
reproduced by theoretical models of galaxy formation. 





We concentrate on four Semi-Analytical Models (SAMs): 
Menci et al. (2008), Merson et al. (2012) 
(based on the SAM of Bower et al. 2006), 
Somerville et al. (2012), and Lu et al.  (specifically, in the best fit case of the
 rendition described in Lu et al., 2013; see also Lu et al. 2011, 2012).

These models vary both in the way they assemble the dark matter halos as well as in
the prescriptions adopted for the various physical mechanisms involved in galaxy
formation. Here we briefly mention the physical processes relevant in the discussion
of our result. We refer the reader to the original papers for full details of the
models, and outline here only the major differences.  In the Menci model the merging histories of dark matter haloes are
described through Monte Carlo simulations, the Merson model uses dark matter merger
histories extracted from the Millenium simulation (Springel et al. 2005), while the
Somerville and Lu models use the Bolshoi N-body simulations (see Klypin et al. 2011
for details).  We note that the
resolution of these N-body simulations in practice sets a lower limit to the mass
(hence luminosity) of the smaller galaxies traced in the simulations, and hence to
the depth of the luminosity distributions that we are presenting.
Different modes for the suppression of gas cooling in massive halos are
implemented in the SAMs, 
either by artificially turning off the cooling of hot gas in haloes above a tunable
mass threshold (as implemented in the Lu model), or 
 through the inclusion of a specific model for AGN feedback. The latter can be
either related to the active AGN phase 
(quasar mode), or ignited by the continuous accretion of gas in dark matter haloes
undergoing
quasi-hydrostatic cooling (radio mode). The quasar mode is implemented in both the
Somerville and in the Menci model, 
which both assume the AGN activity to be  triggered by galaxy interactions: the
latter can be constituted only by 
 galaxy merging (as in the Somerville model) or include also  galaxy fly-by (as in
the Menci SAM). 
The radio mode AGN feedback is implemented in the Merson and in the Somerville
model, and 
influences galaxies with massive black holes at the center of massive haloes.  \\

  All these models provide simulated galaxy samples, for which magnitudes in any
desired filter set are given. The Somerville and Lu  models provide lightcone mock
catalogs that mimic the geometry of the UDS and GOODS-S fields, while the other two average over larger areas of the sky. 
Galaxy magnitudes are computed from the predicted
star formation and chemical enrichment histories using single stellar population
model (SSP).  All these models use the SSP model of Bruzual \& Charlot (2003) but
they assume a different initial mass function (IMF): Kennicutt IMF (Merson),
Salpeter IMF (Menci), Chabrier IMF (Somerville and Lu).  We do not correct the luminosities and colours for the different
IMF because the parameters of each model have been tuned to reproduce observations.

The dust extinction
affecting the above magnitudes is computed from the dust optical depth and
applying the appropriate attenuation to the luminosity at various wavelengths. The only exception is the Lu et al. model that does not include dust extinction.

For all these models  we have extracted the
$BzK$ magnitudes and applied exactly the same $BzK$ criterion we applied to real
data, as described in the previous sections. 
  
\begin{figure}
  \centering
 \includegraphics[width=9cm]{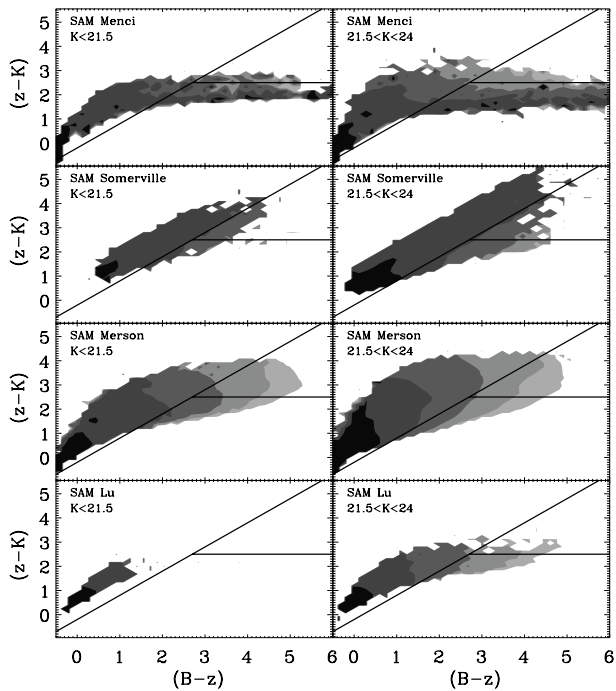}
   \caption{$B$-$z$ versus $z$-$K$ colour-colour diagram for model galaxies in two 
bins of $K$-magnitude. The filled contours correspond to the predicted average
values of the specific star formation rate  of model galaxies for bins of
different $B$-$z$ and $z$-$K$. The SSFR values are equally spaced in logarithmic
scale from SSFR=10$^{-13}$yr$^{-1}$ for the lightest filled region to
SSFR=10$^{-9}$yr$^{-1}$ for the darkest.}
     
     \label{sam}
  \end{figure}

 First of all we have tested whether these models are able to reproduce the passive galaxies 
selected by the $pBzK$ criterion. To this aim, Figure \ref{sam} shows the distribution of model galaxies at $1.4<z<2.5$ in the $BzK$ plane 
in two bins of $K$ magnitude, one corresponding to $K$ magnitudes below the observed peak in the number counts at $K\simeq 21.5$, the other 
corresponding to fainter magnitudes. The gray-scale shows the average specific SFR (SSFR) at a given position in the 
$BzK$ plane, in decades ranging from quiescent galaxies (SSFR=10$^{-13}$yr$^{-1}$) to actively star-forming (SSFR=10$^{-9}$yr$^{-1}$). All models reproduce the distribution of galaxies in the $BzK$ diagram reasonably well. On average, the SSFR of galaxies populating the upper right region of the $B$-$z$ versus $z$-$K$ colour plane is lower than it is in those populating the $BzK$ star forming region ($BzK\ge$-0.2). In the case of the Menci model there is a somewhat large fraction of objects slightly outside the $BzK$ region, but this is still consistent with the expected color 
evolution of galaxies at various ages in the redshift range $1.4<z_{phot}<2.5$ (Daddi et al. 2004). 
The Lu et al. model is devoid of bright $sBzK$ galaxies in the reddest region ($(B-z)\simeq (z-K) \simeq 2$); this region is mostly populated 
by dusty star-burst galaxies, and hence this deficit is expected, considering that the Lu et al. model does not 
include a prescription for dust absorption.  In particular, in all models, the $pBzK$ criterion selects galaxies with SSFR$<$10$^{11}$yr$^{-1}$ 
which is well below the inverse of the Hubble time at $z=2$ 
($1/t_{Hubble}$=3$\times$10$^{-10}$yr$^{-1}$), thus the selected $pBzK$ galaxies are actually passive (see e.g. Franx et al. 2008; 
Fontana et al.  2009).

 In any case, most models predict that some fraction of the population of passive galaxies at $z\simeq 2$ is actually located outside the classical $pBzK$ region, such that present observations might be 
missing a whole class of objects in the color-color plane. As shown in Daddi et al. (2004), and in
Grazian et al. (2007), the $pBzK$ selection criterion preferentially
selects galaxies that have been passively evolving for about 1\,Gyr: 
lowering the threshold in ($z-K$) corresponds to introducing passive galaxies
that are progressively younger, and hence become more numerous. Hence,
we should expect that lowering this threshold would also increase the
number of observed galaxies. Unfortunately this
cannot be easily checked in the data, since lowering
the threshold in ($z-K$) immediately introduces also galaxies at lower
redshift that contaminate the selection.

\begin{figure}
  \centering
  \includegraphics[width=9cm]{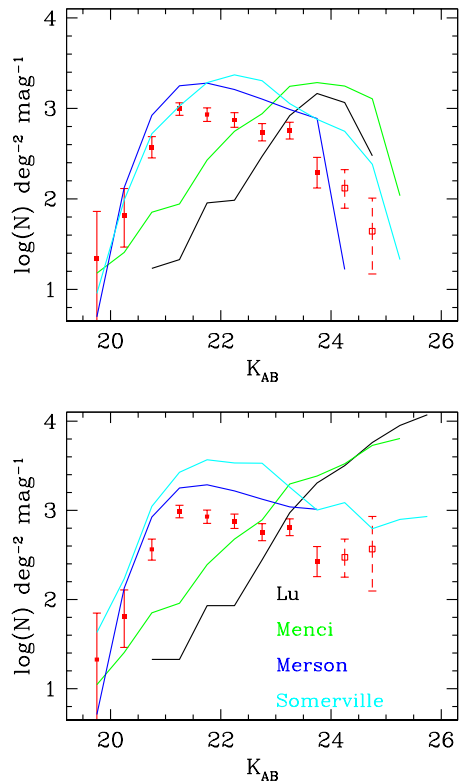}
   \caption{Number counts of $pBzK$ galaxies compared with model prediction of Merson et al. (2012) in blue, Menci et al. (2008) in green, Lu et al. (2013)
in black, and Somervile et al. (2012) in cyan.
{\it Upper panel}. Mock catalogs from theoretical models (solid colored lines) are convolved with the same photometric errors and incompleteness that affect the data. Points refer to the raw observed counts,  without any correction for incompleteness.
{\it Lower panel}. Here the theoretical models are shown without any correction, while observed data are corrected for incompleteness as described in the text.
       }
     \label{sim}
  \end{figure}

This allows us to turn back to our central result, concerning the decrease in the number density of passive galaxies at faint luminosities.
We compare in  Figure~\ref{sim}  the predicted and observed  number counts of $pBzK$
galaxies as a function of $K$ magnitude. 
Since the observations of $pBzK$ galaxies are severely affected by incompleteness, we decided to follow a twofold approach.
In the upper panel we plot the comparison between the observed data and the theoretical data, including the effect of incompleteness in the theoretical data only: the catalog produced by the theoretical models has 
been convolved with the instrumental noise as in the simulations used for estimating the completeness. In the lower panel we have plotted the data corrected for incompleteness and the theoretical models with no correction. As expected the two approaches provide us with a consistent picture, with the latter giving us also the opportunity to show what the models predict at magnitudes fainter than our survey.

We first note that two  models, Merson et al. (2012) and
Somerville et al. (2012),  exhibit a global trend that is {\it qualitatively}
consistent with the observed break in the luminosity distributions. 
Their predicted number density reaches a peak at $K \simeq 21.5$ similar to what we
observe in the real data, and
decreases at fainter magnitudes. 
However, they overpredict  the absolute number density of the passive galaxies  at
faint magnitudes. 
By contrast, the Menci
and the Lu models perform poorly both at large and at small masses. They underestimate the number of massive 
passive galaxies,  do not even predict a turn-over. In both cases the number counts 
continue to increase exponentially towards fainter magnitudes
(i.e. to lower stellar masses).

These discrepancies highlight that some of the prescriptions adopted to describe the main baryonic processes lead to noticeable differences among the various models and in some case need a substancial revision. We remark that our criterion is effective to select galaxies that are undergoing a passive phase since at least $\simeq 1$~Gyr, already at $z\simeq 2$: this allows us to test the implementation of  physical processes that not only quench the star--formation process but also prevents it from starting again after a relatively short time.
In general terms,  the formation  of passive galaxies depends on several proceess, like: the efficiency of gas cooling and star--formation rate, the ejection of gas out of galaxies (due for instance by AGN and stellar feedback) and processes that deplete their gas reservoir (like stripping of cold gas in halo merging). A higher efficiency in the star--formation process and a lower feedback lead to a fast transformation of the gas content of a given halo into stars: in the absence of further cooling or in the presence of stripping of hot gas during halo merging, this processes makes galaxies quiescent. On the contrary, lower efficiencies and higher feedback with subsequent cooling eventually lead to much more extended (albeit intermittent) epochs of star-formation, making galaxies bluer. While these trends are common to all models, their specific implementation varies among them, leading to the differences that we  observe.

At the bright magnitudes the differences
in model predictions can be interpreted in terms of the AGN feedback, that is often
identified as a potentially major factor in the quenching of star-formation
activity, and is suspected to be more effective in massive galaxies.
Based on this comparison, we can speculate that  the radio mode feedback seems to be
more efficient in producing  bright passive galaxies  than the QSO mode and halo
quenching model.

At the faint magnitudes ($K\lesssim23$)  all models over-estimate the number of
passive objects. This is directly related to the well-known tendency of hierarchical
clustering scenarios  to produce steep luminosity functions (see e.g. Somerville \&
Primack 1999; Cole et al. 2000; Menci et al. 2002; Benson et al. 2003). 
Hierarchical
clustering scenarios predict the high-redshift
progenitor galaxies to be characterized by low virial temperature
and high densities, making the gas cooling extremely efficient at $z\gtrsim 3$; the
short dynamical time scales and the
frequent merging events rapidly convert such available cold gas into stars. In the absence of competing effects, the
rapid star formation in low-mass galaxies at $z\gtrsim 3$ would result in an
excess of red low-mass galaxies already at $z\simeq 2$, where our survey is sensitive.

Another physical process that is very important in predicting the number of faint red galaxies is the stripping  of the hot gas reservoir of galaxies residing in  subhaloes. 
All models, indeed, assume that when a halo merges with a larger halo (becoming a subhalo) its hot gas content is stripped away. The galaxy residing in the sub-halo rapidly consumes its cold gas reservoir and thereafter becomes red and passively evolving. This is particularly effective  in dense environments at intermediate and high redshift.  The current implementation 
of such a mechanism produces an excess of red faint galaxies in all theoretical models developed so far. This aspect was explored at low redshift by Weinmann et al. 2006, comparing a SAM model to the observations from the SDSS survey. Guo et al. 2011 modified the implementation of the process in the model in order to delay the stripping of hot gas after halo merging. Even in this case, however, the excess of faint red galaxies is not completely removed. Other, more exotic possibilities have been also explored: for instance, Nierenberg et al. 2013 have shown that, by adopting a Warm Dark Matter power spectrum, the number of satellite galaxies is significantly reduced.

There is no doubt that this excess is at least partly responsible for the excess of predicted galaxies that we observe at low masses in the $pBzK$ population. For this reason Merson et al. (2012) made the test of removing the satellite galaxies altogether from the predicted counts of $pBzK$, finding a good agreement with the observed counts. This is clearly an indication that this process is involved in the overestimate. 
In the case of the Menci and Lu models the discrepancy at faint luminosities is much larger, despite the fact that gas stripping is included in a very similar fashion, suggesting that other processes are also involved.
Unfortunately, it is impossible to make a more accurate test of the relative contribution of central and satellite galaxies, since some fraction of the observed data are also satellite galaxies: given that extended large scale structures are detected in both our fields in the redshift range of $BzK$ galaxies, this fraction could be substantial. This remains a clearly important test to be executed that deserves a dedicated analysis of the data (both in field surveys as well as in high redshift clusters).




%
 %

\section{Summary and Discussion}
In this paper we have  exploited new deep, wide-field 
$K$-band imaging, performed with the High Acuity Wide field $K$-band Imager
(HAWK-I) on VLT as part of the HAWK-I UDS and GOODS-S survey (HUGS;
VLT Large Program), to study the population of passive evolving
galaxies at $z\sim2$.  Crucially, both survey fields
possess $B$ and $z$-band imaging, as well as a wealth of 
other multi-wavelength data, extending from the $U$-band 
to the mid-infrared {\it Spitzer}, and containing 
deep HST imaging assembled as part of the CANDELS HST Treasury program.

To define a sample of star forming and quiescent galaxies at redshift
between $1.4<z<2.5$, we use the $BzK$ criterion proposed by Daddi et
al. (2004).  Thanks to the depth
of our observations, we have now extended the selection to magnitudes fainter
than was possible in previous analysis.  At bright magnitudes the
observed number counts of the $sBzK$ and $pBzK$ galaxies are in good agreement
with previous studies, in particular with McCracken et al. (2010)
as shown in Figure~\ref{RawCounts}. 

In particular we focus on passively evolving galaxies, i.e. the so called $pBzK$ galaxies. 
We find a total of 101 $pBzK$ galaxies in the UDS CANDELS field, and 89
in the GOODS-S field. We have demonstrated through simulations that -- in our deep sample -- the
statistics of passive galaxies down to $K\simeq 22.5$ are not significantly affected by
incompleteness, and that the latter is still treatable down to $K\simeq 24$. Thanks to the depth of our observations we are now able 
to place on a secure footing the earlier results on the $pBzK$ number counts
previously reported by Hartley et al. (2008) and
McCracken et al. (2010).  

Our central result is that the number counts of $pBzK$ galaxies show a flattening at $K_{s}\sim21$, with indications of a turn-over at $K_{s}>22$, equivalent to rest-frame absolute $I$-band magnitudes of $M_{I}=-23, -22$ respectively.  Converted into stellar mass, our result corresponds to a decrease in the number density of passively-evolving galaxies at stellar masses below $10^{10.8}\,{\rm M_{\odot}}$ for a Salpeter IMF.  As judged against the still steeply-rising number counts of the overall galaxy population at these redshifts, this turnover is fairly abrupt, indicating that at high redshift the mechanism that quenches star-formation activity is much less efficient below this mass limit.

{ Another central result is related to the redshift distribution of  $pBzK$  galaxies. As expected, nearly all $pBzK$ galaxies at $K<23$ are at $1.4<z<2.5$. However, at $K>23$ a sharp transition appears to exist, with most of the $pBzK$  galaxies having a redshift  definitely above $z=2.5$, typically at $2.7<z<3.4$. This result confirms the existence of passively-evolving 
galaxies at $z>2.5$, as suggested by a number of recent papers (e.g. Fontana et al. 2009; Straatman et al. 2014). }

We have compared our observed number counts with the predictions of several 
semi-analytical models of galaxy formation and evolution,
in particular with the models of Menci et al. (2008), Somerville et al. (2012),  Merson et al. (2013),
and Lu et. al. (2011).  Among these SAMs only two, 
(Somerville et al. (2012) and Merson et al. (2013))  {\it qualitatively}
predict the shape of the number counts, showing a
turn-over at a stellar mass close to that observed in the data, but  
they do not reproduce the absolute observed density of the passive galaxies.
In contrast, the other two models fail to show this turn-over, and predict an exponential increase 
of passive objects to faint magnitudes. 
This comparison suggests that the distribution of number density (with magnitude or 
stellar mass) of quiescent galaxies at these redshifts offers a critical test for hierarchical models,
and can place strong constraints on the detailed baryonic physical processes involved in galaxy 
formation and evolution. 

The importance of this basic finding should not be 
underestimated.  Indeed,  current models of galaxy formation 
are able to provide acceptable fits to the observed properties of the
{\it bulk} of the galaxy population, such as luminosity functions  or
color distributions. What we show here is that the
physical mechanisms that they implement are not able to   correctly reproduce the {\it
extreme} star formation histories 
characterizing the $pBzK$ galaxy population. 

The observed discrepancies raise pressing questions concerning the comparison between observations and 
the current models of galaxy formation in a cosmological context.
First, are the current implementations of the star-formation quenching process (tuned to match the observed local 
luminosity functions) effective in reproducing the evolution of galaxies in the color-color plane? 
While at large masses the most likely candidate is the AGN feedback, at small masses stellar feedback and gas stripping in satellite galaxies are probably the main processes that produce the observed abundance of the $pBzK$ population. They both need to be better understood and parametrized to provide us with a satisfactory fit to the observed number counts.

Second, is the relative role of the different star-formation modes correctly implemented in the models?
In particular, it is known that both merging and disk instabilities can provide starbursts which add to the 
secular conversion of gas into stars (the "quiescent mode" of star formation). Current galaxy-formation models 
are characterized by different implementations of three modes of star formation; while the bulk of the galaxy population could 
be less dramatically affected by the relative role of such processes, the abundance of $pBzK$ galaxies characterized by 
extremely early star-formation histories could be extremely sensitive to the relative importance of the 
different star-formation modes.

 Unfortunately, it is currently not possible to determine the specific features
that, within each model, produce the observed trends on the quenching
of star formation.  This requires a more detailed investigation that
we plan to explore in  a forthcoming paper.

\begin{acknowledgements} We thank the referee David Wilman for a very useful report. The authors are grateful to the ESO staff for their support during the Hawk-I observations.  JSD acknowledges the support of the Royal Society via a Wolfson Research Merit award, and also the support of the European Research Council via the award of an Advanced Grant.  RJM acknowledges the support of the European Research Council via the award of a Consolidator grant.  AF and JSD acknowledge the contribution of the EC FP7 SPACE project ASTRODEEP (Ref.No: 312725).  This work is based in part on observations (program GO-12060) made with the NASA/ESA {\it Hubble Space Telescope}, which is operated by the Association of Universities for Research in Astronomy, Inc, under NASA contract NAS5-26555.  This work is also based in part on observations made with the {\it Spitzer Space Telescope}, which is operated by the Jet Propulsion Laboratory, California Institute of Technology under NASA contract 1407.  
Program based in part on the data collected in the ESO programs 60.A-9284,
181.A0717, LP 186.A-0898 and 085.A-0961. 
\end{acknowledgements}

{}

\end{document}